\documentclass[cameraready]{Interspeech}

\usepackage{amsmath,amssymb,amsfonts}
\usepackage{algorithmic}
\usepackage{graphicx}
\usepackage{hyperref}
\usepackage{textcomp}
\usepackage{xcolor}
\usepackage{kotex}
\usepackage{pifont}
\usepackage{booktabs}
\usepackage{multirow}
\usepackage{nth}

\title{Whisper-CD: Accurate Long-Form Speech Recognition \\using Multi-Negative Contrastive Decoding}

\author[affiliation={1}, orcid=0009-0002-1809-5362]{Hoseong}{Ahn}
\author[affiliation={1}, orcid=0009-0003-8039-2293]{Jeongyun}{Chae}
\author[affiliation={1}, orcid=0009-0006-0935-1781]{Yoonji}{Park}
\author[affiliation={1}, orcid=0000-0002-0123-3100]{Kyuhong}{Shim}

\address{
    $^1$ Sungkyunkwan University, Republic of Korea
}

\email{ \{hoseong8115, jyunchae, yoonji4024, khshim\}@skku.edu }


\usepackage{comment}

\usepackage{url}
\begin{document}
\maketitle

\begin{abstract}
Long-form speech recognition with large encoder--decoder models such as Whisper often exhibit hallucinations, repetition loops, and content omissions.
These errors can accumulate and be further amplified when the previous segment's transcription is used as decoding context.
We propose \textbf{Whisper-CD}, a training-free contrastive decoding framework that contrasts clean-audio logits against negative logits computed from three acoustically motivated perturbations: Gaussian noise injection, silence signal, and audio temporal shift.
We aggregate these negatives via the log-sum-exp operator, building a unified multi-negative objective for token-by-token decoding.
Across five English long-form benchmarks, Whisper-CD reduces WER by up to 24.3\,pp on CORAAL and shows 48\% faster token generation throughput than beam search.
Because Whisper-CD operates purely at inference time, it can be applied as a drop-in replacement to already-deployed Whisper systems without retraining.
\end{abstract}

\section{Introduction}\label{sec:intro}

Large-scale encoder--decoder models, such as Whisper~\cite{radford2023whisper}, have significantly advanced automatic speech recognition (ASR), yet long-form transcription remains error-prone.
When such models process recordings with prolonged silences, acoustic degradation, or distribution shifts, they often produce fluent but unsupported text.
This phenomenon is broadly termed \emph{hallucination}~\cite{koenecke2024careless, baranski2025investigation,atwany2025lost,park2025evaluating}.
Such errors are difficult to detect because the model typically predicts them with high confidence.
Moreover, long-form audio is usually processed in a divide-and-conquer manner, which can propagate errors as decoding proceeds across segments.

In practice, hallucination in long-form ASR appears in three common patterns: (i)~silence-region hallucination, where the model generates fictitious words during non-speech intervals; (ii)~repetition loops that carry over between segments; and (iii)~content omissions, where parts of the spoken content are dropped.
Architectural approaches for long-form processing~\cite{le2025chunkformer, gong2023longfnt} improve segmentation and context utilization, but do not directly target these error patterns.
Hallucination-specific mitigation techniques, such as VAD-based chunking, constrained decoding, and fine-tuning of hallucinatory attention heads~\cite{wang2025calm}, typically focus on a single error pattern while introducing additional components or requiring model retraining.
This motivates a unified decoding-time method that can address diverse error patterns without parameter updates.

Contrastive decoding~\cite{li2023contrastive} (CD) improves generation quality and has been shown to reduce hallucinations in vision-language~\cite{leng2024mitigating} and natural language processing tasks, including machine translation~\cite{chuang2024dola,sennrich2024mitigating,waldendorf2024contrastive,zhu2025alleviating}.
By contrasting logits from a target generation process against logits from a negative process that amplifies undesirable behavior, CD steers token selection away from incorrect outputs without modifying model parameters.
For example, VCD~\cite{leng2024mitigating} corrupts the input image with heavy noise to obtain negative logits that reflect weaker visual evidence, so tokens preferred under this condition are down-weighted during decoding.
Analogously, the negative process can be instantiated through audio perturbations that reduce speech evidence; to the best of our knowledge, we are the first to apply contrastive decoding to ASR.

In this paper, we propose \textbf{Whisper-CD}, a training-free \textbf{C}ontrastive \textbf{D}ecoding framework for long-form ASR.
Whisper-CD suppresses hallucinated generation by contrasting clean audio logits against multiple negative logits.
Whisper-CD leverages three audio-specific negative signals: (i)~Gaussian noise injection, (ii)~silence-only input (all-zero waveform), and (iii)~temporal shift of the input waveform.
These perturbations are designed to capture common and distinct long-form ASR failure modes.
The three negative logits are combined using the log-sum-exp operator with a single uniform contrastive coefficient~$\alpha$, producing a unified multi-negative CD objective that addresses all three failure patterns.
Experiments on five long-form ASR benchmarks show that Whisper-CD consistently reduces word error rate (WER) and hallucinations compared to the baseline while adding substantially less computational overhead than beam search.

\begin{figure*}[t]
    \centering
    \includegraphics[width=1.0\linewidth]{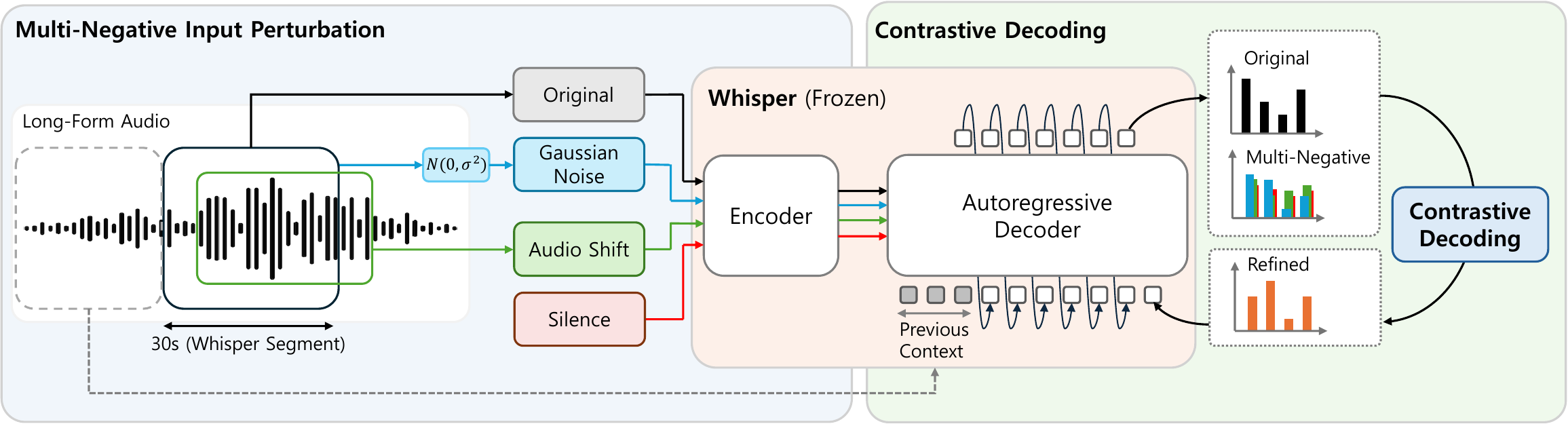}
    \caption{Overview of the proposed Whisper-CD. Each audio segment is processed through four parallel paths, comprising the original signal and three acoustically perturbed variants (Gaussian noise injection, silence signal, and audio temporal shift). Each path produces decoder logits conditioned on the corresponding encoder output, and contrastive decoding steers token selection away from hallucinated outputs such as repetition loops and content omissions.}
    \label{fig:overview}
\end{figure*}

\section{Related Work}\label{sec:related}

\subsection{Challenges in Long-form ASR}
Long-form ASR remains challenging even for large-scale models trained on massive and diverse corpora~\cite{radford2023whisper}.
Existing methods include joint segmentation-and-decoding~\cite{huang2022e2e}, factorized neural transducers~\cite{gong2023longfnt}, and masked chunking conformers~\cite{le2025chunkformer}.
While effective, these approaches generally require architectural changes or retraining, which limits their applicability to already-deployed ASR systems.

\subsection{Hallucinations in Whisper ASR}
Whisper is known to hallucinate frequently on non-speech segments, noisy audio, or out-of-distribution inputs~\cite{koenecke2024careless, baranski2025investigation}, making hallucination a serious concern in long-form transcription.
A representative failure case is the so-called \emph{bag of hallucinations}~\cite{baranski2025investigation}, a set of stock phrases (e.g., ``Thank you for watching'') that the model generates with high confidence when no genuine speech is present~\cite{koenecke2024careless, baranski2025investigation}.
Several mitigation strategies have been proposed, including fine-tuning hallucinatory attention heads for non-speech segments~\cite{wang2025calm} and timestamp-aware training for more verbatim transcription~\cite{wagner2024crisperwhisper}.
In contrast, training-free decoding-time approaches that can address diverse hallucination patterns remain underexplored; in particular, contrastive decoding has not yet been applied to reduce ASR hallucination.

\subsection{Test-time Improvement in ASR}
A complementary line of research seeks to improve ASR quality at inference time without retraining the model.
Classical examples include beam search, which explores multiple decoding hypotheses, and language model rescoring, which re-ranks $n$-best lists using an external language model.
More recently, generative error correction (GER) leverages LLMs to post-correct ASR transcripts~\cite{yang2023generative, hu2024listen}, and test-time adaptation methods instead update model parameters per utterance via entropy minimization~\cite{lin2022listen} or sequential-level generalized entropy minimization~\cite{kim2023sgem}.
However, these approaches have fundamental limitations.
Beam search broadens the decoding space, but does not alter the model's token distribution; when hallucinated tokens receive high probability, increasing the beam size can still concentrate the probability mass onto the same incorrect transcript.
GER operates after the initial decoding pass, and thus cannot directly influence token selection during decoding.
In contrast, our method directly adjusts the token distribution at inference time.
\section{Whisper-CD}\label{sec:method}

\begin{table*}[t]
    \caption{Speech recognition results of Whisper-CD across different Whisper model variants.}
    \vspace{-0.2cm}
    \label{tab:main_results}
    \renewcommand{\arraystretch}{1.1}
    \setlength{\tabcolsep}{10pt}
    \centering
    \resizebox{\linewidth}{!}{%
        \begin{tabular}{ll ccccc cc}
            \toprule
            \multirow{2}{*}{\textbf{Model}} & \multirow{2}{*}{\textbf{Method}} & \multicolumn{5}{c}{\textbf{WER} (\%)$\downarrow$} & \multicolumn{2}{c}{\textbf{Throughput}} \\
            \cmidrule(lr){3-7} \cmidrule(lr){8-9}
            &                                & CORAAL         & VoxPopuli      & Earnings22     & TED-LIUM       & REV-16         & Speed (tokens/s)$\uparrow$ & RTF$\downarrow$    \\
            \midrule
            \multirow{2}{*}{Large-v3}       & Baseline                       & 208.76         & 44.95          & 520.94         & 66.42          & 173.69         & 30.6             & 0.2886 \\
            & + \textbf{CD}                   & \textbf{45.77} & \textbf{19.86} & \textbf{57.08} & \textbf{25.62} & \textbf{21.38} & 27.3             & 0.1655 \\
            \midrule
            \multirow{2}{*}{Large-v3-Turbo} & Baseline                       & 38.75          & 30.63          & 33.25          & 12.93          & 19.82          & 168.9            & 0.0239 \\
            & + \textbf{CD}                   & \textbf{14.43} & \textbf{25.71} & \textbf{16.16} & \textbf{10.11} & \textbf{14.81} & 144.2 & 0.0346 \\
            \bottomrule
        \end{tabular}
    }
\end{table*}

\subsection{Motivation}
Whisper processes long-form audio by splitting the input into 30-second segments and decoding each.
The previous segment's transcription can optionally be provided as context for consistency across segments (see Figure~\ref{fig:overview}).
Counterintuitively, Whisper's performance often degrades when the previous context is given. 
This behavior is well known in practice, as the official documentation%
\footnote{\url{https://github.com/openai/whisper/blob/7858aa/whisper/transcribe.py\#L278}}%
recommends disabling context passing when performance is unsatisfactory.

We further quantify this effect: On Whisper Large-v3, conditioning on previous context \textit{increases} WER by over 190\,pp on CORAAL~\cite{quartey2020coraal_vld} and over 500\,pp on Earnings22~\cite{del2022earnings22}.
This degradation is mainly driven by error accumulation.
Once an erroneous transcription (e.g., hallucinated text and repetition loops) is passed as context, it may bias the current segment's decoding and propagate errors to subsequent segments.
Such hallucinations are difficult to recover from, even with beam search, and are more prominent in larger decoders.
Rather than disabling context passing, which sacrifices the model's inherent ability to utilize long-range cues, we seek to preserve \textit{context-aware decoding} while suppressing errors.
This motivates a logit-level decoding strategy based on contrastive decoding, which down-weights tokens that remain highly probable under deliberately degraded audio conditions.

\subsection{Contrastive Decoding for ASR}
Contrastive decoding steers token selection by contrasting the model logits under a target (positive) generation process with those under a negative process.
We adapt this principle to ASR by deriving the negative signal from \textit{acoustically perturbed} versions of the input waveform, rather than from a weaker model.
Let $\mathbf{x}$ denote an input waveform and $g(\cdot)$ a perturbation function that produces $\tilde{\mathbf{x}} = g(\mathbf{x})$.
At decoding step $t$, given the previously generated tokens $y_{<t}$, we compute logits using two forward passes of the same ASR model:
\begin{equation}
    \ell_t^{\text{pos}} = f_{\theta}\big(\textbf{x}, y_{<t}\big), \quad \ell_t^{\text{neg}} = f_{\theta}\big(\tilde{\textbf{x}}, y_{<t}\big)
\end{equation}
where $f_{\theta}(\cdot)$ outputs a logit vector over the vocabulary.
The contrastive logits are then defined as:
\begin{equation}
    \ell_t^{\text{CD}} = (1 + \alpha)\, \ell_t^{\text{pos}} \,-\, \alpha\, \ell_t^{\text{neg}}
    \label{eq:cd}
\end{equation}
where $\alpha > 0$ controls the contrastive strength.
Token selection is performed from $\ell_{t}^{\text{CD}}$.
This procedure is training-free: it requires no parameter updates and can be applied to any token-by-token generation ASR model at inference time.

\subsection{Perturbation Strategies}
We design three perturbation functions to instantiate a negative generation process for contrastive decoding.
The key idea is to deliberately reduce or distort acoustic evidence so that the resulting logits reflect the model's biases or tendencies rather than the true speech content.

\subsubsection{Gaussian Noise Injection}
The input waveform $\mathbf{x}$ is corrupted with additive Gaussian noise calibrated to a target signal-to-noise ratio (SNR).
Specifically, the corrupted waveform is $\tilde{\mathbf{x}} = \mathbf{x} + \boldsymbol{\epsilon}$, where $\boldsymbol{\epsilon} \sim \mathcal{N}(\mathbf{0}, \sigma^2 \mathbf{I})$ and $\sigma$ is chosen to achieve the target SNR.
This acoustic corruption makes local phonetic evidence unreliable while retaining the coarse acoustic structure, producing negative logits that represent the model's behavior under reduced acoustic clarity.

\subsubsection{Silence Signal}
The input spectrogram is set to all zeros, which removes spectral structure entirely.
Under this condition, the decoder behaves as if it is continuing the text with minimal acoustic evidence, revealing its unconditional textual prior.
The resulting logits tend to emphasize silence-region hallucination patterns, including the ``bag of hallucinations'' phrases reported in prior analyses~\cite{koenecke2024careless, baranski2025investigation}.

\subsubsection{Audio Temporal Shift}
The input waveform is shifted leftward by $\Delta_s$ seconds to create a controlled misalignment between acoustic content and the segment's temporal position.
Concretely, the first $\Delta_s$ seconds of audio are discarded, and the same duration is zero-padded at the end, so the decoder receives future audio content earlier than expected.
This temporal mismatch disrupts the alignment between the decoder's prefix context and the local acoustics, producing negative logits that represent segment-boundary failure tendencies.

\subsection{Multi-Negative Contrastive Decoding}
To simultaneously address multiple long-form failure patterns, we combine $K$ negative signals ($K{=}3$) into a single decoding step.
We aggregate the negative logits using a log-sum-exp operator with a temperature $\tau$:
\begin{equation}
    \ell_t^{\text{CD}} = \left(1 + \alpha\tau\right) \ell_t^{\text{pos}} \,-\, \alpha\tau\, \log\Big(\frac{1}{K} \sum_{k=1}^{K} \exp\big(\ell_{k,t}^{\text{neg}} \,\,/\,\, \tau\big)\Big).
    \label{eq:bcd}
\end{equation}
where $\ell_{k,t}^{\text{neg}}$ denotes the negative logits obtained from the $k$-th perturbed input $\tilde{\mathbf{x}}_k$.
Setting $K{=}1$ and $\tau{=}1$ recovers Eq.~\eqref{eq:cd}.
A small $\tau$ concentrates influence on the dominant negative (approximating a max), while a large $\tau$ approaches a smoother average.
We set $\tau{=}1.0$, corresponding to the log-sum-exp aggregation, and vary $\alpha \in [0.5, 2.0]$.

\subsection{Inference}\label{ssec:inference}
For efficiency, we compute encoder outputs for the clean input and all $K$ perturbed inputs in a single batched forward pass.
During autoregressive decoding, we compute all paths in parallel by packing them along the batch dimension, requiring only a single batched decoder forward pass per step while reusing the same prefix tokens $y_{<t}$.
Since Whisper-CD intervenes only at the logit level, it preserves the model's existing capabilities, such as language identification and timestamp generation, and remains compatible with standard inference pipelines, including beam search and token-level constraints.

\section{Experimental Results}\label{sec:experiment}

\begin{figure*}[t]
    \centering
    \includegraphics[width=1.0\linewidth]{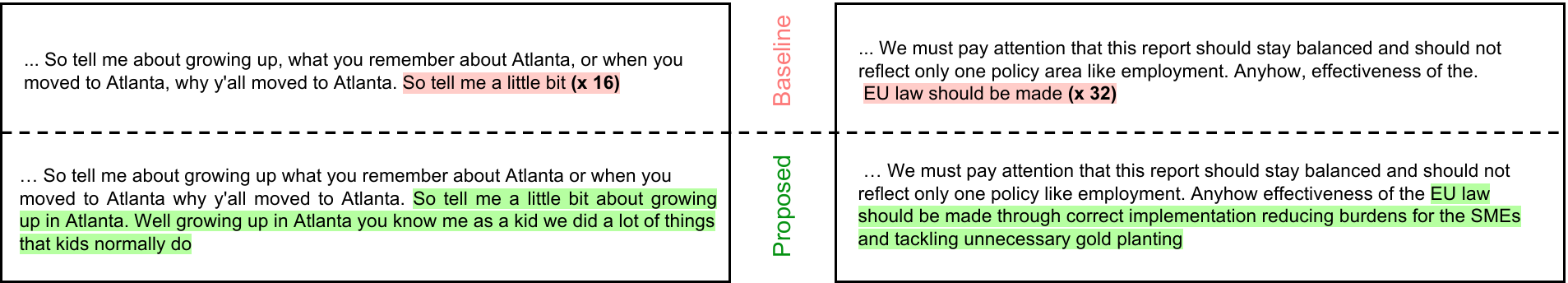}
    \caption{Qualitative examples on the same audio inputs. The baseline falls into repetition loops (\textcolor{red}{red}), while Whisper-CD breaks the loop and recovers the correct transcription (\textcolor{green!60!black}{green}).}
    \label{fig:qualitative}
    \vspace{-0.1cm}
\end{figure*}

\subsection{Setup}\label{ssec:experiment_setup}

\noindent\textbf{Datasets.}
We evaluate on five English long-form benchmarks covering diverse speaking styles and acoustic conditions.
CORAAL~\cite{quartey2020coraal_vld} uses 14 recordings (11.5\,h) from the VLD region.
Earnings22~\cite{del2022earnings22} is a 13-file subset (14.3\,h) from the full 125-file corpus ($\sim$120\,h), selected at regular duration intervals (every \nth{3} file when sorted by duration) to ensure uniform coverage.
VoxPopuli~\cite{wang2021voxpopuli,fox2024updated} is a 50-file subset (1.4\,h) from the long-form reconstitution of the test split, filtered for recordings between 60 and 300 seconds and sampled at every \nth{9} file.
TED-LIUM~\cite{fox2024updated, hernandez2018ted} uses the full long-form reconstitution test set (11 files, 3.1\,h).
REV-16~\cite{radford2023whisper} uses 16 recordings (16.1\,h).

\vspace{0.15cm}\noindent\textbf{Inference Setup.}
We employ Whisper's long-form pipeline with 30-second segments and previous-context passing enabled, matching the default configuration.
We use greedy decoding unless otherwise specified and disable the temperature fallback mechanism to isolate the effect of contrastive decoding.

\vspace{0.15cm}\noindent\textbf{Models.}
We conduct experiments on two Whisper model sizes: Large-v3 and Large-v3-Turbo.
Both variants reuse the same weights for all negative paths, which are computed jointly in a single batched forward pass (Section~\ref{ssec:inference}).
All analysis experiments are conducted on Whisper Large-v3-Turbo.

\vspace{0.15cm}\noindent\textbf{Perturbation hyperparameters.}
We set $\text{SNR}_{\text{dB}}=10$ for Gaussian noise injection.
Audio temporal shift applies a leftward shift of $\Delta_s = 7$\,s.

\vspace{0.15cm}\noindent\textbf{Metrics.}
We report word error rate (WER, \%$\downarrow$) as the primary metric.
To assess computational cost, we additionally report decoding throughput in tokens-per-second and real-time factor (RTF), measured on a single NVIDIA A100 80GB PCIe with a batch size of~1.

\subsection{ASR Performance and Inference Efficiency}\label{ssec:main_results}
Table~\ref{tab:main_results} shows that Whisper-CD reduces WER across all five long-form ASR benchmarks.
Note that the Large-v3 baseline WER exceeds 100\% on several datasets because repetition loops inflate the output length well beyond the reference.
CD suppresses these repetitions, though the WER remains above Turbo's baseline; we discuss this gap in Section~\ref{ssec:analysis}.
Figure~\ref{fig:qualitative} shows qualitative examples of how CD eliminates repetition loops and non-speech hallucinations.

Regarding throughput, CD introduces three additional decoding paths, yet the reduction in total generated tokens can offset this cost.
For Large-v3, eliminating repetition loops lowers overall wall-clock time, improving RTF over the baseline.
For Large-v3-Turbo, the additional paths incur a modest slowdown that remains substantially faster than beam search (Table~\ref{tab:beam_search}).

\subsection{Analysis}\label{ssec:analysis}

\vspace{0.15cm}\noindent\textbf{Impact of $\alpha$.}
Table~\ref{tab:ablation_alpha} shows the effect of the contrastive strength~$\alpha$.
Datasets with higher baseline WER generally benefit from stronger contrastive signals, while the cleaner TED-LIUM is sensitive to over-subtraction and degrades beyond $\alpha{=}1.5$.
Notably, all nonzero $\alpha$ values improve over the baseline on CORAAL and Earnings22, indicating that CD is broadly robust for hallucination-prone speech recordings.

\begin{table}[t]
    \caption{Ablation on contrastive decoding strength ($\alpha$). Setting $\alpha=0$ corresponds to the baseline without CD.}
    \vspace{-0.2cm}
    \label{tab:ablation_alpha}
    \centering
    \renewcommand{\arraystretch}{1.}
    \setlength{\tabcolsep}{17pt}
    \resizebox{\columnwidth}{!}{%
        \begin{tabular}{lccc}
            \toprule
            \multirow{2}{*}{$\alpha$} & \multicolumn{3}{c}{\textbf{WER}(\%)$\downarrow$} \\
            \cmidrule(lr){2-4}
            & CORAAL         & Earnings22     & TED-LIUM       \\
            \midrule
            0.0 & 38.75          & 33.25          & 12.93          \\
            \cmidrule(lr){1-4}
            0.5 & 26.58          & \textbf{16.16} & 11.19          \\
            1.0 & \textbf{14.43} & 17.70          & 11.65          \\
            1.5 & 19.65          & 19.44          & \textbf{10.11} \\
            2.0 & 20.82          & 21.77          & 11.00          \\
            \bottomrule
        \end{tabular}%
    }
    \vspace{-0.1cm}
\end{table}

\vspace{0.15cm}\noindent\textbf{Individual strategies.}
Table~\ref{tab:ablation} compares perturbation strategies by using each as the sole negative.
No single strategy consistently improves performance across all datasets; for example, silence degrades TED-LIUM to 21.62\% despite improving CORAAL.
In contrast, the multi-negative combination (Table~\ref{tab:main_results}) achieves lower WER than the single-strategy variants on most datasets, suggesting that the proposed aggregation effectively leverages complementary negative signals.

\begin{table}[t]
    \caption{Ablation on three audio perturbation strategies.}
    \vspace{-0.2cm}
    \label{tab:ablation}
    \centering
    \renewcommand{\arraystretch}{1.1}
    \setlength{\tabcolsep}{13pt}
    \resizebox{\columnwidth}{!}{%
        \begin{tabular}{lccc}
            \toprule
            \multirow{2}{*}{\textbf{Strategy}} & \multicolumn{3}{c}{\textbf{WER}(\%)$\downarrow$} \\
            \cmidrule(lr){2-4}
            & CORAAL   & Earnings22 & TED-LIUM \\
            \midrule
            Gaussian    & 38.11    & 19.50      & 12.49    \\
            Silence     & 18.99    & 17.41      & 21.62    \\
            Audio Shift & 18.77    & 15.54      & 13.81    \\
            \bottomrule
        \end{tabular}%
    }
    \vspace{-0.2cm}
\end{table}

\vspace{0.15cm}\noindent\textbf{Effect of model scale.}
Despite its larger capacity, Large-v3 yields substantially higher WER under CD than Large-v3-Turbo (45.77\% vs.\ 14.43\% on CORAAL).
We hypothesize that this gap stems from differences in failure severity and decoder-driven error propagation.
In particular, Large-v3 often enters deep repetition loops that can push the baseline WER above 200\%.
Once such loops are established, logit-level contrast alone may be insufficient to consistently steer decoding back to a correct trajectory, because the model assigns overwhelming probability mass to self-reinforcing continuations.
Moreover, when a previous-segment context is provided as a prefix, a stronger decoder may rely more heavily on that textual context, which can amplify the impact of earlier mistakes.

\vspace{0.2cm}\noindent\textbf{Comparison to beam search.}
Table~\ref{tab:beam_search} compares CD with beam search (beam size~5).
Beam search improves performance on CORAAL but degrades TED-LIUM, whereas CD achieves lower WER on both datasets with higher throughput, offering a more favorable accuracy-throughput trade-off.

\begin{table}[t]
    \caption{Comparison with beam search (beam size = 5).}
    \vspace{-0.2cm}
    \label{tab:beam_search}
    \centering
    \renewcommand{\arraystretch}{1.}
    \setlength{\tabcolsep}{8pt}
    \resizebox{\columnwidth}{!}{%
        \begin{tabular}{lcccc}
            \toprule
            \multirow{2}{*}{\textbf{Method}} & \multicolumn{2}{c}{\textbf{WER}(\%)$\downarrow$} & \multicolumn{2}{c}{\textbf{Throughput}} \\
            \cmidrule(lr){2-5}
            & CORAAL & TED-LIUM & Speed$\uparrow$ & RTF$\downarrow$ \\
            \midrule
            Baseline      & 38.75  & 12.93    & 174.3           & 0.0246          \\
            + Beam Search & 22.65  & 17.50    & 99.0            & 0.0436          \\
            + \textbf{CD} & 14.43  & 10.11    & 147.0           & 0.0302          \\
            \bottomrule
        \end{tabular}%
    }
    \vspace{-0.2cm}
\end{table}

\subsection{Discussion and Future Work}\label{ssec:discussion}

\noindent\textbf{Dynamic $\alpha$.}
Since the optimal $\alpha$ differs across datasets and model sizes (Tables~\ref{tab:main_results} and~\ref{tab:ablation_alpha}), predicting and adjusting $\alpha$ dynamically per segment or per token would improve robustness on diverse acoustic conditions.

\vspace{0.15cm}\noindent\textbf{Additional perturbations.}
Our multi-negative framework is not restricted to a fixed set of negatives; other audio transformations, such as frequency masking, chunk shuffling, or temporal warping, could potentially serve as additional negative signals.

\vspace{0.15cm}\noindent\textbf{Decoder-only ASR models.}
Recent decoder-only ASR models~\cite{abouelenin2025phi4,liu2025voxtral} process audio and text in a single stream, making it less straightforward to inject perturbed audio while preserving the text prefix.
Adapting the negative path construction to such architectures is a natural next step, building on the precedent of visual contrastive decoding in vision-language models.

\section{Conclusion}\label{sec:conclusion}
We proposed Whisper-CD, a training-free contrastive decoding framework for long-form speech recognition.
Whisper-CD contrasts clean-audio logits against three acoustically motivated negative signals: Gaussian noise injection, silence signal, and audio temporal shift.
This multi-negative formulation mitigates long-form failure patterns, including non-speech hallucinations, repetition loops, and content omissions without any parameter update.
Experiments on five English long-form ASR benchmarks show consistent WER reductions across model configurations, with up to a 24.3\,pp improvement on CORAAL, while adding only modest computational overhead relative to greedy decoding and remaining substantially faster than beam search.



\section{Acknowledgments}
This work was supported by Institute of Information \& communications Technology Planning \& Evaluation (IITP) grant funded by the Korea government (MSIT) (RS-2019-II190421, Artificial Intelligence Graduate School Program (Sungkyunkwan University)).

\section{Generative AI Use Disclosure}
The authors used large language models, specifically ChatGPT (OpenAI) and Claude (Anthropic), to assist with proofreading and improving the clarity and fluency of the manuscript.

\bibliographystyle{IEEEtran}
\bibliography{reference,datasets}

\end{document}